\documentclass[11pt,a4paper]{article}
\pdfoutput=1

\usepackage{jcappub}
\usepackage{graphicx}
\usepackage{epsfig}
\usepackage{subfig}
\usepackage{comment}
\usepackage{slashed}
\usepackage{soul}
\usepackage{tabularx}
\usepackage{physics}
\usepackage[normalem]{ulem}
\usepackage{afterpage}
\usepackage{placeins}
\usepackage{booktabs}
\usepackage{float}

\newcommand{\mc}{\mathcal}

\usepackage{array}
\newcolumntype{C}[1]{>{\centering\let\newline\\\arraybackslash\hspace{0pt}}m{#1}}

\newcommand{\be}{\begin{equation}} 
\newcommand{\ee}{\end{equation}}
\newcommand{\bea}{\begin{equation}\begin{aligned}} 
\newcommand{\eea}{\end{aligned}\end{equation}}
\newcommand{\ber}{\begin{eqnarray}}
\newcommand{\ear}{\end{eqnarray}}

\def\lsim{\mathrel{\raise.3ex\hbox{$<$\kern-.75em\lower1ex\hbox{$\sim$}}}}
\def\gsim{\mathrel{\raise.3ex\hbox{$>$\kern-.75em\lower1ex\hbox{$\sim$}}}}

\newcommand{\Mpc}{{\rm Mpc}}

\newcommand{\ie}{{\it i.e.}}
\newcommand{\eg}{{\it e.g.}}

\newcommand{\td}{{\rm d}}

\newcommand{\mpl}{M_{\rm P}}

\newcommand{\epsh}{\epsilon_{H}}

\newcommand{\etah}{\eta_{H}}

\newcommand{\PR}{\mathcal{P}_{\mathcal{R}}}

\definecolor{DigitColor}{rgb}{0.5,0.5,0.5}
\newcommand{\digits}[1]{\textcolor{DigitColor}{#1}}

\definecolor{Red}{rgb}{1,0,0}
\definecolor{Blue}{rgb}{0,0,1}
\definecolor{Green}{rgb}{0,1,0}

\begin{document}

\title{Primordial black holes and inflation from double-well potentials}

\author[a]{Alexandros Karam,}
\author[a]{Niko Koivunen,}
\author[a]{Eemeli Tomberg,}
\author[a]{Antonio Racioppi,}
\author[a]{and Hardi Veerm\"{a}e}

\affiliation[a]{Laboratory of High Energy and Computational Physics, NICPB, \\ 
R{\"a}vala pst.~10, Tallinn, 10143, Estonia}

\emailAdd{alexandros.karam@kbfi.ee}
\emailAdd{niko.koivunen@kbfi.ee}
\emailAdd{eemeli.tomberg@kbfi.ee}
\emailAdd{antonio.racioppi@kbfi.ee}
\emailAdd{hardi.veermae@cern.ch}

\abstract{We investigate the formation of large peaks in the inflationary curvature power spectrum from double-well potentials. In such scenarios, the initial CMB spectrum is created at large field values. Subsequently, the inflaton will cross one of the minima and will decelerate rapidly as it reaches the local maximum at the origin, either falling back or crossing it. During this final phase, a significant peak in the curvature power spectrum can be generated. Our analysis reveals that this class of models produces more pronounced peaks than most quasi-inflection point scenarios with less tuning for the model parameters. Finally, we construct an explicit theoretically motivated inflationary scenario that is consistent with the latest CMB observations and capable of generating sufficiently large curvature perturbations for primordial black holes. 
}

\maketitle

%%%%%%%%%%%%%%%%%%%%%%%%%%%%%%%%%%%%%%%%%%%%%%%%%%%%%%%%%%%%%%
\section{Introduction}
\label{sec:intro}
%%%%%%%%%%%%%%%%%%%%%%%%%%%%%%%%%%%%%%%%%%%%%%%%%%%%%%%%%%%%%%

The inflationary paradigm~\cite{Starobinsky:1980te, Kazanas:1980tx, Sato:1980yn, Guth:1980zm, Linde:1981mu, Albrecht:1982wi, Linde:1983gd} can explain several cosmological conundra as well as provide a quantitative mechanism for producing the observed primordial fluctuations~\cite{Starobinsky:1979ty, Mukhanov:1981xt, Hawking:1982cz, Starobinsky:1982ee, Guth:1982ec, Bardeen:1983qw}. However, beyond the scales probed by the cosmic microwave background (CMB), the primordial curvature power spectrum is essentially unconstrained due to the lack of empirical data and may contain much stronger fluctuations than the ones that have been observed in the CMB.

High peaks in the primordial power spectrum can lead to the production of primordial black holes (PBHs) when the curvature fluctuations re-enter the cosmological horizon in the early Universe~\cite{Hawking:1971ei, Carr:1974nx, Carr:1975qj}. Although the PBH abundance is constrained by various observables (for a review see \eg~\cite{Carr:2020gox}), asteroid mass PBHs remain a viable dark matter (DM) candidate~\cite{Niikura:2017zjd, Katz:2018zrn, Montero-Camacho:2019jte}. They correspond to an enhanced power spectrum at scales around $\mathcal{O}(10^{13})\ \Mpc^{-1}$. Peaked spectra at scales $k < 10^{7}\ \Mpc^{-1}$ are associated with solar mass or heavier PBHs. These may be related to the BH-BH merger events observed in LIGO-Virgo~\cite{LIGOScientific:2016aoc, LIGOScientific:2018mvr, LIGOScientific:2020ibl, LIGOScientific:2021djp} if they make up $\mathcal{O}(0.1\%)$ of DM~\cite{Raidal:2017mfl, Raidal:2018bbj, Vaskonen:2019jpv, Hutsi:2020sol, Wong:2020yig, DeLuca:2020jug, Hall:2020daa, Franciolini:2021tla}. An even smaller fraction of heavier PBHs may have served as seeds of supermassive BHs and cosmic structures~\cite{1975A&A....38....5M, Duechting:2004dk, Kawasaki:2012kn, Clesse:2015wea, Carr:2018rid, Liu:2022bvr, Hutsi:2022fzw}.

Large peaks in the primordial power spectrum can be produced in a vast array of inflationary models, with single-field inflation being the simplest%~\cite{Ivanov:1994pa, Carr:1994ar, Bullock:1996at, Kawasaki:1997ju, Kawasaki:1998vx, Yokoyama:1998pt, Kohri:2007qn, Saito:2008em, Bugaev:2008bi, Drees:2011hb, Drees:2011yz, Kawasaki:2016pql, Garcia-Bellido:2017mdw, Domcke:2017fix, Ezquiaga:2017fvi, Kannike:2017bxn, Germani:2017bcs, Motohashi:2017kbs, Di:2017ndc, Ballesteros:2017fsr, Ballesteros:2020qam, Hertzberg:2017dkh, Cicoli:2018asa, Ozsoy:2018flq, Biagetti:2018pjj, Ezquiaga:2018gbw, Dalianis:2018frf, Gao:2018pvq, Rasanen:2018fom, Ballesteros:2018wlw, Carr:2018poi, Passaglia:2018ixg, Dalianis:2019asr, Atal:2019cdz, Drees:2019xpp, Kuhnel:2019xes, Bhaumik:2019tvl, Fu:2019ttf, Atal:2019erb, Arya:2019wck, Mahbub:2019uhl, Mishra:2019pzq, Ballesteros:2019hus, Ashoorioon:2019xqc, Fu:2019vqc, Fu:2020lob, Nanopoulos:2020nnh, Ozsoy:2020kat, Ragavendra:2020sop, Kefala:2020xsx, Ng:2021hll, Solbi:2021wbo, Inomata:2021uqj, Stamou:2021qdk, Wu:2021zta, Bastero-Gil:2021fac, Dalianis:2021iig, Solbi:2021rse, Zheng:2021vda, Cheng:2021lif, Teimoori:2021pte, Heydari:2021gea, Kawai:2021edk, Gangopadhyay:2021kmf, Rezazadeh:2021clf, Inomata:2021tpx, Heydari:2021qsr, Figueroa:2021zah, Wang:2021kbh, Iacconi:2021ltm, Sarkar:2021tsa, Cai:2021zsp, Cicoli:2022sih, Frolovsky:2022ewg, Cheong:2022gfc, Yi:2022anu, Gu:2022pbo, Correa:2022ngq, Frolovsky:2022qpg, Cai:2022erk, Meng:2022low, Frolovsky:2023hqd}
(for a recent overview of single-field scenarios see, \eg,~\cite{Karam:2022nym, Ozsoy:2023ryl}). In this case, the most widely considered scenarios introduce an inflaton potential that contains a weak maximum, or a quasi-inflection point, which causes the inflaton to slow down considerably in an ultra slow-roll (USR) phase as the field rolls over that feature~\cite{Garcia-Bellido:2017mdw, Kannike:2017bxn, Germani:2017bcs, Ballesteros:2017fsr}. Instead of freezing soon after their Hubble exit, curvature perturbations can undergo super-Hubble growth during such a phase, excited by the strong potential feature. As a result, the curvature power spectrum can be greatly enhanced.

A similar---or even stronger---enhancement of the power spectrum can be achieved in models in which the maximum lies between two global minima, also known as double-well potentials. These models arise naturally in the context of spontaneous symmetry breaking, with the best-known example being the Higgs mechanism of the Standard Model (SM)~\cite{Higgs:by:Englert, Higgs:by:Higgs:1, Higgs:by:Higgs:2, Higgs:by:Guralnik}. In beyond the SM models, such as scalar extensions of the SM, Grand Unified Theories, etc., such potentials appear as a means to generate masses around the scale of new physics. Coleman--Weinberg potentials arising when symmetry is broken via quantum corrections~\cite{Coleman:1973jx} have been among the first models proposed for cosmic inflation~\cite{Linde:1981mu, Albrecht:1982wi}.

Regarding PBH production from double-well potentials, there have been a handful of proposals in the literature over the last three decades~\cite{Yokoyama:1998pt, Saito:2008em, Bugaev:2008bi, Briaud:2023eae}. In these papers, the considered potentials were of the Coleman--Weinberg or Higgs-like nature. The disadvantage of these potentials in their minimal form is that for large field values, they behave as $V \propto \phi^4$ and therefore predict a scalar spectral index value of $n_s < 0.95$ and a tensor-to-scalar ratio value of $r = 0.33$, both of which are excluded by existing observations~\cite{Planck:2018jri, BICEP:2021xfz}. To guarantee a sufficiently low tensor-to-scalar ratio value, one should consider potentials that feature a plateau at high field values.

The inflationary timeline of the models considered in this work consists of the usual phases in single-field models for PBHs: the CMB spectrum is produced during an initial slow-roll (SR) phase, which then transitions into a USR phase causing the inflaton to slow down significantly. The latter is succeeded by another phase of SR or constant roll (CR) inflation~\cite{Karam:2022nym}. For double-well potentials, however, the SR conditions are always maximally violated as the field rolls through a global minimum during the SR to USR transition. This leads to a strongly enhanced curvature power spectrum that is qualitatively quite different from typical quasi-inflection point models: in quasi-inflection point scenarios the enhanced power spectrum can be understood well in terms of the duration of the USR phase as well as the Wands duality~\cite{Wands:1998yp} between the USR and the subsequent CR phases. 
In contrast, in scenarios where the field rolls through the global minimum, the shape of the peak is mostly dictated by the details of the SR to USR transition --- it exhibits enhanced oscillatory spectral features that can be interpreted in terms of a brief period of preheating-like particle production. In this paper, we will discuss the general features of double-well models for inflation and PBHs and construct an explicit scenario consistent with CMB measurements and asteroid mass PBH DM.

This paper is structured as follows. In section~\ref{sec:model_building}, we outline the general theoretical considerations that go into building a phenomenologically feasible model for both  PBHs and CMB observables and propose an explicit double-well model. The implications of our results are discussed in Section~\ref{sec:discussion}. We conclude in Section~\ref{sec:concusions}. Natural units $\hbar=c=1$ with $\mpl=1$ and the $(-+++)$ metric signature are used throughout the paper.

%%%%%%%%%%%%%%%%%%%%%%%%%%%%%%%%%%%%%%%%%%%%%%%%%%%%%%%%%%%%%%
\section{Inflation with double-well potentials}
\label{sec:model_building}
%%%%%%%%%%%%%%%%%%%%%%%%%%%%%%%%%%%%%%%%%%%%%%%%%%%%%%%%%%%%%%

%%%%%%%%%%%%%%%%%%%%%%%%%%%%%%%%%%%%%%%%%%%%%%%%%%%%%%%%%%%%%%
\subsection{Theoretical background}
\label{sec:theory}
%%%%%%%%%%%%%%%%%%%%%%%%%%%%%%%%%%%%%%%%%%%%%%%%%%%%%%%%%%%%%%

In order to construct a double-well model that can accommodate both the CMB observables and produce PBHs, we start from the general Jordan frame action for single-field inflation
\be\label{eq:S}
	\mc{S} = \int{\td^4 x \sqrt{-g} \left[ \frac{1}{2}\Omega(\phi) R - \frac{1}{2} K(\phi) (\partial \phi)^2 - U(\phi) \right]}\, ,
\ee
where $R$ is the Ricci scalar and $U(\phi)$ is the Jordan frame potential. When working with inflationary perturbations it is more convenient to work in the Einstein frame. The Weyl rescaling of the metric $g_{\mu\nu} \to g_{\mu\nu}/\Omega(\phi)$ and the field redefinition $\td \varphi/\td \phi = \sqrt{K/\Omega + (3/2)(\Omega_{,\phi}/\Omega)^2}$ transforms the Jordan frame action~\eqref{eq:S} to the canonical form (see \eg~\cite{Kaiser:1994vs, Bezrukov:2007ep, Galante:2014ifa,Jarv:2016sow})
\be\label{eq:S_canonical}
    \mc{S} = \int{\td^4 x \sqrt{-g} \qty[ \frac{1}{2}R-\frac{1}{2}(\partial \varphi)^2 - V ]}\,,    
%    \qquad V(\varphi) \equiv \frac{U(\phi(\varphi))}{\Omega(\phi(\varphi))^2}
\ee
where $V \equiv U/\Omega^2$ is the Einstein frame potential. 

As a simple realization of symmetry breaking, we will consider models in which the Jordan frame field $\phi$ is a real scalar singlet with a $\mathbb{Z}_2$ symmetry under field reversal $\phi \to -\phi$. This symmetry will carry over to the Einstein frame if the field redefinition satisfies $\varphi(\phi=0) = 0$. To obtain a double-well potential, we will assume that the Einstein frame potential should possess a maximum at the origin $\phi = 0$ so the $\mathbb{Z}_2$ symmetry would be spontaneously broken in the (true) vacuum at which the field acquires a vacuum expectation value $|\phi| = v$ and $V(\pm v)=0$ is the (global) minimum.

In the Einstein frame, the homogeneous FRW background evolves according to
\be \label{eq:bg_eom}
    \ddot{\varphi}+3H\dot{\varphi}+V'(\varphi)=0 \, , \qquad
    3H^2 = \frac{1}{2}\dot{\varphi}^2 + V(\varphi) \, ,
\ee
where $H\equiv\dot{a}/a$ is the Hubble parameter, $a$ is the scale factor describing the expansion of the universe, and a dot indicates derivative with respect to the cosmic time $t$. The background evolution is characterized by the Hubble SR parameters
\be \label{eq:SR_parameters}
    \epsh \equiv -\frac{\dot H}{H^2} =  \frac{\dot{\varphi}^2}{2H^2} \, , \qquad
    \etah \equiv -\frac{\ddot H}{2H \dot H} = -\frac{\ddot \varphi}{H \dot \varphi} \, .
\ee
Inflation corresponds to $\epsh < 1$, and SR inflation to $\epsh \, , \, |\etah| \ll 1$.

Linear quantum fluctuations on top of the homogeneous background are described by the Mukhanov--Sasaki variable $u$ that obeys the Fourier space equations of motion~\cite{Mukhanov:1985rz, Sasaki:1986hm, Mukhanov:1990me}
\be \label{eq:SM_eom}
    u_k'' + \qty(k^2 - \frac{z''}{z})u_k = 0 \, , \qquad
    z \equiv a\frac{\dot \varphi}{H} \, .
\ee
Here $k$ is the wavenumber and a prime denotes a derivative with respect to the conformal time $\tau$, with $\td t = a \td \tau$. The Mukhanov--Sasaki variable gives the curvature power spectrum
\be \label{eq:PR}
    \PR 
    = \frac{k^3}{2 \pi^2}\frac{|u_{k}|^2}{z^2}  \, ,
\ee
which freezes to a time-independent shape by the end of inflation. The frozen power spectrum is the standard tool to describe primordial perturbations, which, at large scales, are imprinted on the CMB. At the CMB pivot scale, $k_*=0.05 \mathrm{Mpc}^{-1}$, observations give~\cite{Planck:2018jri,BICEP:2021xfz}
\be \label{eq:CMB_observables}
    A_s \equiv \PR(k_*) = 2.1 \times 10^{-9} \,, \quad %(2.10 \pm 0.03)\times 10^{-9} \, , \quad
    n_s \equiv 1 + \frac{\td\PR(k_*)}{\td\ln k} = 0.9649 \pm 0.0042 \, , \quad
    r < 0.036 \, .
\ee
Here $r$ is the tensor-to-scalar ratio, describing the strength of primordial gravitational waves. The CMB observations guide model building. For an SR model, the observables in terms of the Hubble SR parameters are given as\footnote{For completeness, we also report the spectral parameters during CR~\cite{Yi:2017mxs}
\bea\label{eq:CMB_CR_predictions}
    \PR &= \frac{\Gamma(3/2-\etah)^2}{2^{1-2\etah}\pi^3}\frac{H^2}{\epsh} \,, \qquad
    %2^{2\nu -3} \left[ \frac{\Gamma \left( \nu \right)}{\Gamma \left( 3/2 \right)} \right]^2 \frac{1}{2 \epsh} \left( \frac{H}{2\pi} \right)^2 \left( 1 - \frac{\epsh}{1 + 2 \etah} \right)^{2\nu - 1} \left( \frac{k}{a H} \right)^{3 - 2\nu} \,, \\
    n_s = 4 - \Bigl\lvert 3 + 2\etah \Bigr\rvert \,,  \qquad
    r = 16 Q \epsh  \,,
\eea
where $Q \equiv 2^{3 - \lvert 3 - 2\etah \rvert} \Gamma \left( 3/2 \right)^2/\Gamma \left(  \lvert 3 - 2 \etah \rvert / 2 \right)^2$ and we neglected next-to-leading order terms in $\epsh$ as they are subleading when $\epsh\ll \etah = \mathcal{O}(1)$.
}
\be \label{eq:CMB_SR_predictions}
     A_s = \frac{H^2}{8\pi^2 \epsh} \, , \qquad
     n_s = 1 + 2\etah - 4\epsh \, , \qquad
     r = 16\epsh \,.
\ee
For CMB, these quantities are to be evaluated when the pivot scale exits the Hubble radius, $k_*=aH$. In typical models, the crossing happens 50--60 $e$-folds before the end of inflation.

%%%%%%%%%%%%%%%%%%%%%%%%%%%%%%%%%%%%%%%%%%%%%%%%%%%%%%%%%%%%%%
\subsection{Peaks in $\PR$ and primordial black holes}
\label{sec:PBH}
%%%%%%%%%%%%%%%%%%%%%%%%%%%%%%%%%%%%%%%%%%%%%%%%%%%%%%%%%%%%%%

For the purposes of our paper, it is sufficient to adapt a simplified order-of-magnitude approach to PBH formation. Critical collapse~\cite{Musco:2004ak, Polnarev:2006aa, Musco:2008hv, Musco:2012au, Musco:2018rwt, Kehagias:2019eil, Musco:2020jjb} of density fluctuations exceeding a critical threshold during the radiation-dominated era in the early Universe yields a PBH whose mass is on average comparable to the horizon mass
\be\label{eq:PBH_mass}
    M_k
    \approx 2.8 \times 10^{18}\ {\rm g} \, \left(\frac{k}{10^{14}\ \Mpc^{-1}} \right)^{-2} \, ,
\ee
while a non-negligible PBH abundance can be produced if the curvature power spectrum contains a peak that is of the order 
\be
    \PR(k) \approx \mathcal{O}(10^{-2})\,.
\ee
We will not attempt to make precise predictions of the PBH abundance and instead aim for constructing scenarios that produce spectral features with such a height. This is justified considering that the height of the peak is quite sensitive to the model parameters and can therefore be tuned by a factor of a few without changing the other predictions, for instance, the CMB power spectrum. This freedom can be used to ``correct" the height in order to obtain any desired PBH abundance, which itself is exponentially sensitive to the height of the power spectrum peak. Moreover, the PBH abundance estimates contain potential theoretical uncertainties related to the shape of the spectrum~\cite{Musco:2018rwt} and to non-Gaussianities, \eg, in the form of exponential tails~\cite{Pattison:2017mbe, Ezquiaga:2019ftu, Figueroa:2020jkf, Figueroa:2021zah, Ferrante:2022mui, Gow:2022jfb, Tomberg:2023kli}.

The position of the peak is roughly determined by the duration of the first phase of SR inflation during which the CMB spectrum is generated. In the double-well models considered here, inflation always briefly stops as the field rolls through the global minimum. We will denote the duration of the first inflationary phase in $e$-folds by $N_{\rm I}$. To produce PBHs, the inflaton must typically slow down during a USR-like phase and enter a second phase of SR (or CR) inflation. The duration of the second phase of inflation will be denoted by $N_{\rm II}$. It includes the duration of the USR phase.

We will focus on models that would produce PBHs in the asteroid mass window $10^{17}-10^{22}\ \rm g$~\cite{Laha:2020ivk, Niikura:2017zjd}, where they can constitute all DM without clashing with existing constraints. This window corresponds to comoving scales in the range $k = (5\times 10^{12} - 5\times 10^{14})\ \Mpc^{-1}$ which exit the Hubble horizon $31-37$ $e$-folds after the CMB pivot scale. Assuming a typical total of 55 $e$-folds of inflation, we thus require an additional $18-24$ $e$-folds between the end of inflation and the creation of the enhanced feature in the power spectrum. All in all, we are looking for models with
\be
    N_{\rm I} \in (31, 37)\,,  \qquad
    N_{\rm II} \approx 55 - N_{\rm I} \in (18, 24)\, .
\ee
The second phase of inflation can, however, be much shorter in the case where the total duration of inflation is small, for instance, due to a low scale of inflation or modified early universe cosmologies. These estimates are affected by uncertainties in the total number of $e$-folds which depend on the details of reheating as well as on other potential non-standard cosmological epochs in the Early Universe. The total number of $e$-folds in the example scenarios considered here vary in the range $N_{\rm I} + N_{\rm II} \in [51, 57]$\footnote{A standard computation of the total number of $e$-folds (see \eg~Eq.~(47) in \cite{Planck:2018jri}) with a flat potential and an instant reheating into the Standard Model d.o.f.s yields $N\approx 55.9$ in our example cases with $r\approx0.015$ (see Table~\ref{tab:benchamrk_pts}). Relaxing these assumptions lets $N$ vary by multiple $e$-folds. For example, $N$ can be lowered by assuming a period of dust-like reheating with an equation of state parameter $w=0$ ($20$ e-folds of reheating yields $N\approx 50.9$), and it can be increased with a period kination, where $w=1$ ($3$ $e$-folds of kination yields $N\approx 57.4$).}.

%%%%%%%%%%%%%%%%%%%%%%%%%%%%%%%%%%%%%%%%%%%%%%%%%%%%%%%%%%%%%%
\subsection{An explicit model}
\label{sec:model}
%%%%%%%%%%%%%%%%%%%%%%%%%%%%%%%%%%%%%%%%%%%%%%%%%%%%%%%%%%%%%%

Before proposing an explicit model, let us consider the general criteria it must satisfy. We are looking for double-well potentials that generate the CMB at large field values and PBHs around the origin.

To guarantee consistency with the CMB, potentials with a plateau at large field values are preferred. A simple way to flatten a double-well potential with chaotic behaviour for large field values is to assume that the inflaton field is non-minimally coupled to gravity and/or it has a non-canonical kinetic term of the $\alpha$-attractor type~\cite{Galante:2014ifa, Kallosh:2015zsa}. In such a case, the transformation to the Einstein frame will result in a potential that is asymptotically flat for large positive and negative field values, while retaining the local maximum at the origin and the two global minima on each side of it. 

In a setup like this, the field first experiences SR on one of the plateaus for $N_{\rm I}$ $e$-folds. Since $V(\pm v) = 0$, inflation is always interrupted when the field crosses the potential minimum. Afterward, field evolution can proceed in three different ways: 
\begin{enumerate}
    \item \textbf{Roll-back}: The field does not have enough energy to reach the origin and it turns around. Inflation starts again shortly after the field has passed the closest minimum; this second stage of inflation starts in USR and transitions into its dual SR/CR phase. Inflation ends before the field returns to the minimum.
    \item \textbf{Roll-over}: The inflaton has enough energy to cross the hilltop, but it slows down so much that inflation restarts near the origin. The field crosses the potential hill in USR inflation and transitions into a dual SR/CR phase as it rolls down. Inflation stops again before the field reaches the second minimum. 
    \item \textbf{Fast roll-over}: With a lot of energy, the inflaton quickly rolls over the hilltop at $\phi=0$, not causing the universe to inflate. It starts to oscillate in the second minimum on the opposite side and the universe reheats.
\end{enumerate}
Production of PBHs is possible only in the first two cases in which inflation restarts at the end of a USR period and lasts for $N_{\rm II}$ $e$-folds. On the boundary between these cases, the classical field equations allow the inflaton to stop completely on top of the hill, corresponding to the $N_{\rm II} \to \infty$ limit. Tuning the model parameters close to this critical point lets us tailor the number of $e$-folds spent near the origin. This period comprises both the USR phase and its dual SR/CR phase as was recently discussed in Ref.~\cite{Karam:2022nym} (see also~\cite{Wands:1998yp}). 

\begin{figure*}
    \centering
    \includegraphics[width=0.9\textwidth]{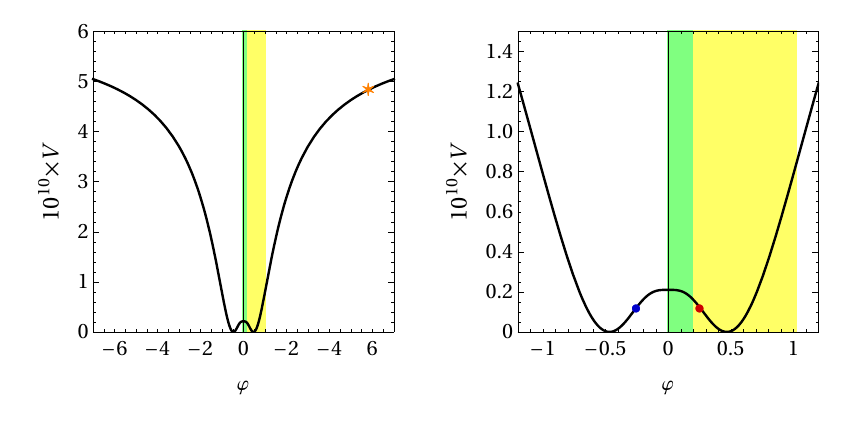}
    \caption{\emph{Left panel:} Einstein frame potential corresponding to parameter values $\lambda= 1.52537\times 10^{-8}$, $\xi=0.01$, $v=0.3851983$ and $\alpha=2v^2/3$. The value of $v$ is tuned to be close to the roll-over/roll-back boundary and corresponds to a roll-over scenario. The green and yellow bands correspond to the USR and temporarily halted inflation, respectively. The orange star corresponds to the CMB scale. \emph{Right panel:} Zoomed-in potential near the origin which represents two cases: roll-over and roll-back. The blue and red dots are the end of the second phase of inflation for roll-over and roll-back, respectively. Note that while the potential and blue dot corresponds to $v=0.3851983$, the red dot corresponds to $v=0.3854377$. The shape of the potential between these two different parameter values is indistinguishable by eye, and we present them in the same figure for ease of illustration.}
    \label{fig:potential}
\end{figure*}

The in-between period, when inflation momentarily ceases, is expected to be negligible in duration in most cases. However, as we will show shortly, with double-well potentials, the peak in the curvature spectrum depends on the features of this SR to USR transition and fluctuations can be enhanced by effects similar to parametric resonance during preheating.

During the last SR/CR phase, one can approximate the Hubble SR parameters with the potential ones and apply Eq.~\eqref{eq:CMB_CR_predictions} to estimate the power spectrum.\footnote{In principle, the duality between the USR and SR/CR allows us to extrapolate the spectral features computed in the SR/CR into the preceding USR phase. However, this is not needed here as the USR has only a mild impact on the full spectrum.} For this, we look at the behaviour of the inflaton close to the origin at $\phi = 0$. Given a $\mathbb{Z}_2$ symmetry, the action~\eqref{eq:S} is expected to depend on $\phi$ through only $\phi^2$ and thus the linear terms vanish in the small-field expansion, for which we assume the form  $\Omega = 1 + \xi \phi^2+ \mathcal{O}(\phi^4)$, $K = 1 + \mathcal{O}(\phi^2)$, and $U = U_0 - m^2 \phi^2/2 + \mathcal{O}(\phi^4)$. Working in the Einstein frame, the relevant quantity in these models is the second SR parameter during the second inflationary phase,
\be\label{eq:eta0}
    \eta_{H,\rm II} \approx \left.\frac{V''}{V} \right|_{\phi=0}
    %\approx \left[\frac{U''}{U} - 2 \Omega''\right]_{\phi=0} %+ \mathcal{O}(\phi^2)
    = - \frac{m^2}{U_0} - 4 \xi \, ,
\ee
which determines the slope of the spectrum. Although $\etah \gtrsim 1$ is theoretically allowed during CR, an extended second inflationary phase $N_{\rm II} \approx \mathcal{O}(20)$ typically requires $\etah \lesssim 1$. This implies $\xi \lesssim 1$ and $m^2 \lesssim U_0$. The last inequality is especially constraining for a tree-level Higgs mechanism, $U = \lambda (\phi^2 - v^2)^2/4$, for which $m^2/U_0 = 2/v^2$ indicating the need for super-Planckian vacuum expectation values ($v \gtrsim 1$) for reasonably small values of $\etah$.\footnote{We tested this claim explicitly by computing the curvature power spectra for a non-minimally coupled Higgs-like field with $U = \lambda (\phi^2 - v^2)^2/4$ and found that PBH production together with a consistent CMB was not possible in any region of the parameter space.}

\begin{figure*}
    \centering
    \includegraphics[width=0.95\textwidth]{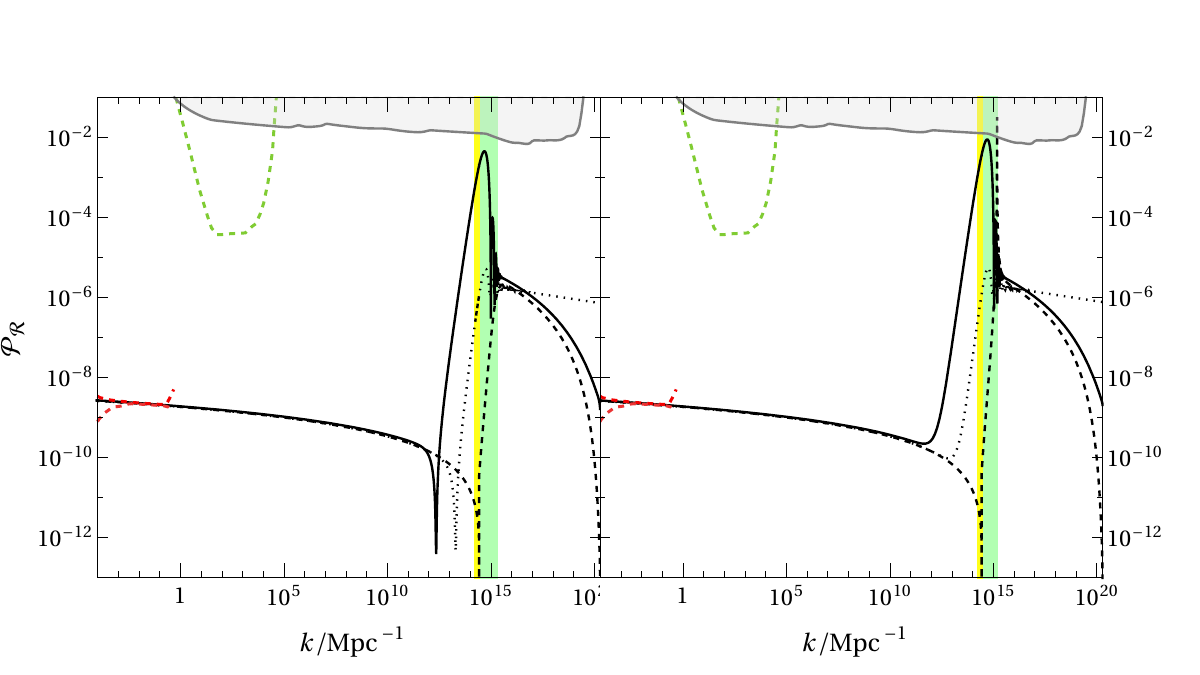}
    \caption{Example power spectra in the cases of  roll-over (left) with $v=0.3851462$ and roll-back (right) with  $v=0.3854781$. In both cases $v$ is tuned to produce $N_{\rm II}=14$, and $n_s = 0.96072 (1\sigma)$, $\lambda=1.79222\times 10^{-8}$ and $\alpha=2v^2/3$. The solid  curve is the exact numerical curvature power spectrum. The dashed and dotted curves show power spectra computed using the SR approximation~\eqref{eq:CMB_SR_predictions}, and the analytical approximation developed for quasi-inflection point models \eqref{eq:PR_approximation}, respectively.
    The green band shows the comoving wavenumbers of the modes that exited during the USR epoch, while the yellow band indicates modes that temporarily re-enter the horizon when inflation is halted and exit again when inflation re-starts. The gray line shows the constraints from the PBH abundance assuming Gaussian curvature fluctuations~\cite{Karam:2022nym}.
    The red dashed line shows CMB observations~\cite{Planck:2018jri} and the green dashed line denotes the constraint from $\mu$-distortions~\cite{Chluba:2012we}.
    } 
    \label{fig:example_spectra}
\end{figure*}

Based on the above reasoning, we will not rely on a non-minimal coupling to generate a plateau at large field values and assume that the symmetry breaking is due to quantum corrections, thus avoiding a large mass around the origin. We will consider a general class of non-minimally coupled $\alpha$-attractor models with a Coleman--Weinberg-like potential~\cite{Linde:1981mu, Albrecht:1982wi}, characterized by
\be \label{eq:action_functions}
    \Omega(\phi) = 1 + \xi \phi^2 \, , \quad
    K(\phi) = \qty(1-\frac{\phi ^2}{6 \alpha })^{-n} \, , \quad
    U(\phi) = \frac{\lambda}{4} \left[ \frac{1}{2}\phi^4 \left(\ln\frac{\phi^2}{v^2}- \frac{1}{2}\right) +  \frac{v^4}{4} \right] \, .
\ee
Near $\phi=0$, the Coleman--Weinberg-like potential $U$ introduces a double well with minima at $\phi=\pm v$. Near $|\phi| = \sqrt{6\alpha}$, the non-canonical kinetic factor $K$ makes the derivative $\td\varphi/\td\phi$ diverge, so the Einstein frame potential $V(\varphi)$ develops plateaus suitable for SR inflation. The same mechanism is used in $\alpha$-attractor models with $n=2$~\cite{Galante:2014ifa, Kallosh:2015zsa}. However, to obtain a sufficiently high $n_s$ at the CMB scales, we will choose $n=4$. Thus, our ansatz \eqref{eq:action_functions}, especially the non-standard exponent $n=4$, is mostly motivated by phenomenology and we will not provide a UV framework for this set-up. We require $v < \sqrt{6\alpha}$ to retain both the double-well and the plateau. The non-minimal coupling $\Omega$ tunes the relative heights of the hilltop and the plateaus, but, as we will show later, its contribution is rather minor for scenarios in which PBHs can be produced. An example of the Einstein frame potential for the model~\eqref{eq:action_functions} is plotted in Fig.~\ref{fig:potential}. 

To study the cosmological predictions of the model in the interesting cases of roll-over and roll-back, we solve the background field equations~\eqref{eq:bg_eom} numerically, starting from parameter values $\xi$, $\alpha$, $\lambda$, and $v$. We then solve the Mukhanov--Sasaki equation~\eqref{eq:SM_eom} for multiple $k$-modes to obtain the power spectrum (see Appendix~\ref{appendix:numerics} for details). In all of our examples, we set the CMB pivot scale on the plateau at a point that produces a desirable $n_s$ according to~\eqref{eq:CMB_SR_predictions}. We then fix $\lambda$ by requiring the correct power spectrum normalization $A_s$ at this scale. We tune $v$ to fix the duration of the second inflationary phase $N_{\rm II}$.

\begin{figure*}
    \centering
    \includegraphics[width=1.0\textwidth]{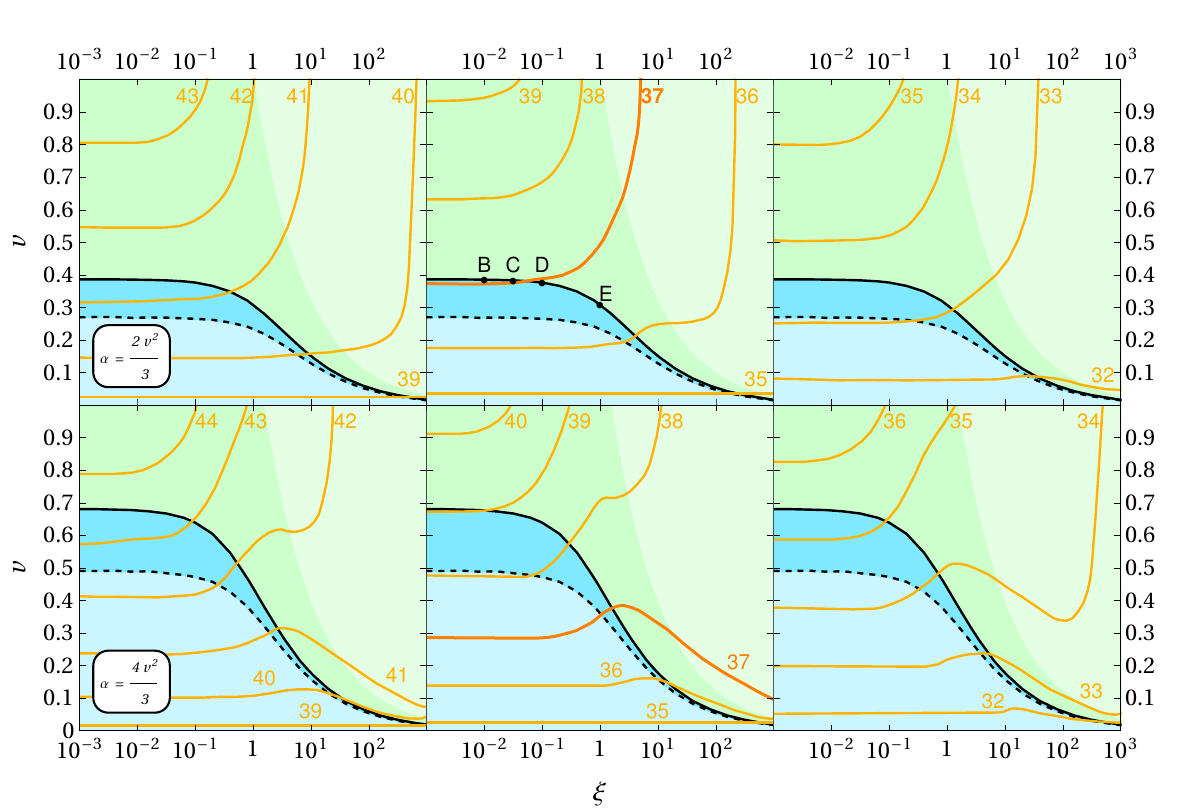}
    \caption{A scan over the $(\xi,v)$-space with $\alpha=2v^2/3$ \emph{(upper panels)} and $\alpha=4v^2/3$ \emph{(lower panels)}. Both green-shaded regions stand for roll-back; in the lighter green region, the hilltop is higher than the plateau, preventing roll-over by a simple energy conservation argument. The blue region stands for roll-over and lighter blue for fast roll-over. PBH production can take place on the black curve between the roll-back and roll-over regions. The orange curves give the number of $e$-folds of the first phase of inflation, starting from the CMB. The three panels correspond to three values of $n_s$: from left to right, the central value $n_s =0.9649$, the $1\sigma$ value $n_s=0.96072$, and the $2\sigma$ value $n_s=0.95652$.  The black dots correspond to benchmark points in Table~\ref{tab:benchamrk_pts}.}
    \label{fig:scan}
\end{figure*}

\begin{table}
\begin{center}
\begin{tabular}{ccccccccc}
\toprule
 & $\xi$ &  & $v$ & $r$ &  $N_{\rm I}(1\sigma)$ & $\mathcal{P}_{\mathcal{R},{\rm max}}$ %& $k_{\rm max}$ & $M_{PBH}/ 10^{17} {\rm g}$ 
\\
\midrule
A & $0$ & RO & $0.386\digits{2076}$ & $0.01573$ & $37.0$ & $1.7\times10^{-2}$ %& $5.5\times10^{14}$ & $0.93$ 
\\
& & RB & $0.386\digits{5523}$ & $0.01574$ & $37.0$ & $3.5\times10^{-2}$ %& $5.8\times10^{14}$ & $0.83$
\\
B & $10^{-2}$ & RO & $0.385\digits{1983}$ & $0.01574$ & $37.0$ & $1.5\times10^{-2}$ %& $5.5\times10^{14}$ & $0.93$
\\
&  & RB & $0.385\digits{4377}$ & $0.01563$ & $37.0$ & $2.4\times10^{-2}$ %& $5.8\times10^{14}$ & $0.83$
\\
C & $10^{-1.5}$ & RO & $0.38\digits{299285}$ & $0.01544$ & $37.0$ & $2.4\times10^{-2}$ %& $5.5\times10^{14}$& $0.94$
\\
& & RB & $0.38\digits{30859}$ & $0.01544$ & $37.0$ & $3.0\times10^{-2}$ %& $5.5\times10^{14}$& $0.93$
\\
D & $10^{-1}$ & RO & $0.3760\digits{49595}$ & $0.01490$ & $36.9$ & $3.7$ %& $5.4\times10^{14}$& $0.94$
\\
& & RB & $0.3760\digits{524}$ & $0.01490$ & $36.9$ & $3.7$ %& $5.4\times10^{14}$& $0.94$
\\
E & $1$ & RO & $0.30842795518270935781\digits{585}$ & $0.01091820$ & $36.5$ & $1.7\times10^{29}$ %& $5.1\times10^{14}$& $0.11$
\\
& & RB & $0.30842795518270935781\digits{915}$ & $0.0109182097$ & $36.5$ & $5.9\times10^{28}$ %& $5.1\times10^{14}$& $0.11$
\\
\bottomrule
\end{tabular}

\end{center}

\caption{Benchmark points for $\alpha = 2v^2/3$ for roll-over (RO) and roll-back (RB), respectively. The $\mathcal{P}_{\mathcal{R},{\rm max}}$ is the maximum value in the peak of the power spectrum and $k_{\rm max}$ is the location of the peak. The duration of the first phase of inflation, $N_{\rm I}$, is determined by demanding that $n_s=0.96072$, which lies $1\sigma$ below the central value. The vacuum expectation value $v$ is tuned to produce $N_{\rm II}\approx 19$. The last digits in $v$ differing in the roll-over and roll-back cases are highlighted in gray. At all points, we obtain $M_{k} \approx 10^{17}\ \rm g$ and $k_{\rm max} \approx 5 \times 10^{14}\ {\rm Mpc}^{-1}$.}
\label{tab:benchamrk_pts}
\end{table}

Figure~\ref{fig:example_spectra} shows example power spectra in the roll-over and roll-back cases. After the low-$k$ SR plateau, the roll-over spectrum dips down, analogously to the widely studied quasi-inflection point models~\cite{Karam:2022nym}. Interestingly, the roll-back spectrum shows no such dip, which is the main difference between the two cases. Around the intermediate non-inflationary period, the power spectrum then grows in both cases, spurred on by the violent background behaviour --- the momentary end of inflation and the subsequent USR and CR phases. A high peak is followed by a number of oscillations and a longer plateau.
%Quasi-inflection point models may also produce such oscillations, but only when the transition between the SR and USR phases is artificially sharp~\cite{Karam:2022nym}. Here, the brief intermediate period during which inflation stops produces a naturally sharp SR to USR transition, enhancing the first peak so that it dominates over the rest of the spectrum and is responsible for all PBH formation.

Figure~\ref{fig:scan} shows scans over the parameter space for $\alpha=2v^2/3$ and $\alpha=4v^2/3$ and varied $\xi$ and $v$. The spectral tilt $n_s$ at the CMB scales is varied between the three panels from the central value in the first panel to values lower by $1\sigma$ and $2\sigma$, in the second and third panels, respectively. The coloured regions correspond to the different behaviours 1--3, with the blue shading showing models in which the inflaton can roll over the maximum and the green shading denoting when the field rolls back. Increasing $\xi$ decreases the height of the plateau. In the light green region, the plateau becomes lower than the hilltop, and only the roll-back case is possible. Decreasing $v$ lowers the local maximum at $\phi=0$ and in the light-blue region, the field can roll over it without ever initiating a second phase of inflation. PBHs can be produced in a fine-tuned region along the black line that separates the roll-back and roll-over regions. 

Fixing $N_{\rm II}\approx 19$ $e$-folds, singles out one roll-back and one roll-over point for a given $\xi$. Table~\ref{tab:benchamrk_pts} lists such benchmark points, with $n_s$ chosen to lie $1\sigma$ below the central value. We see that small values of $\xi$ produce the desirable $\PR\sim10^{-2}$, with the horizon mass at the edge of the asteroid mass window. For low $\xi \lesssim 0.1$, the background evolution, as well as the curvature power spectrum, is nearly independent of $\xi$ and thus matches roughly the one for minimally coupled models. On the other hand, a significant enhancement in $\PR$ can be observed when $\xi \gtrsim 0.1$. This is partly because, by Eq.~\eqref{eq:eta0}, $\eta_{H,\rm II} \approx -4\xi$ and the spectrum has a large slope also during the last inflationary CR period, $\ie$, after the peak. For a fixed $N_{\rm II}$, this can further enhance the height of the peak and lead to unphysically large fluctuations as exemplified by case E in Table~\ref{tab:benchamrk_pts}.

PBHs may be produced in the asteroid mass window when $N_{\rm I} \lesssim 37$. Models with a lower $n_s$ allow for a shorter initial period of inflation, as can be seen from Fig.~\ref{fig:scan}. For the central value $n_s =0.9649$, we see that $N_{\rm I} > 37$ for all of the parameter space and thus the produced PBHs would be too light and evaporate. For $n_s = 0.9607$, $1\sigma$ below the central value, we observe that $N_{\rm I} > 37$ is attainable for a sufficiently large $\xi$. The lower bound on $\xi$ depends on $\alpha$. With $n_s$ lying $2\sigma$ below the central value, $N_{\rm I} < 37$ is realized in the entire parameter space we considered. Moreover, as in quasi-inflection point models, accounting for the running of $n_s$ can permit lower values of $n_s$, thus relaxing the tension with the CMB observations~\cite{Planck:2018jri, Ballesteros:2020qam}.

We note that low values of $n_s$ are a typical problem in PBH plateau models, see \eg~\cite{Ballesteros:2017fsr, Rasanen:2018fom, Dalianis:2018frf, Mahbub:2019uhl, Iacconi:2021ltm, Frolovsky:2022qpg}: $\alpha$-attractor models predict the correct $n_s$ when the main phase of inflation lasts for $50$ $e$-folds and reducing this to our $30$ throws these predictions off. In equation~\eqref{eq:action_functions}, we combated this by choosing the $K$-exponent $n=4$ instead of the conventional $n=2$, making $n_s$ larger. With this modification, we found workable solutions within the $2\sigma$ uncertainty bounds, if not with the central value.

The results are qualitatively similar for both of the $\alpha$ values we considered. Lowering $\alpha/v^2$ while keeping $\xi$ fixed raises the peak of $\PR$, and below $\alpha=2v^2/3$, asteroid-mass PBHs are always overproduced. On the other hand, for $\alpha=4v^2/3$, $\xi$ needs to be larger than in the case of $\alpha=2v^2/3$ to obtain the same $n_s$ and $N_{\rm II}$, indicating a larger $\PR$ peak.

\begin{figure*}
    \centering
    \includegraphics[width=0.85\textwidth]{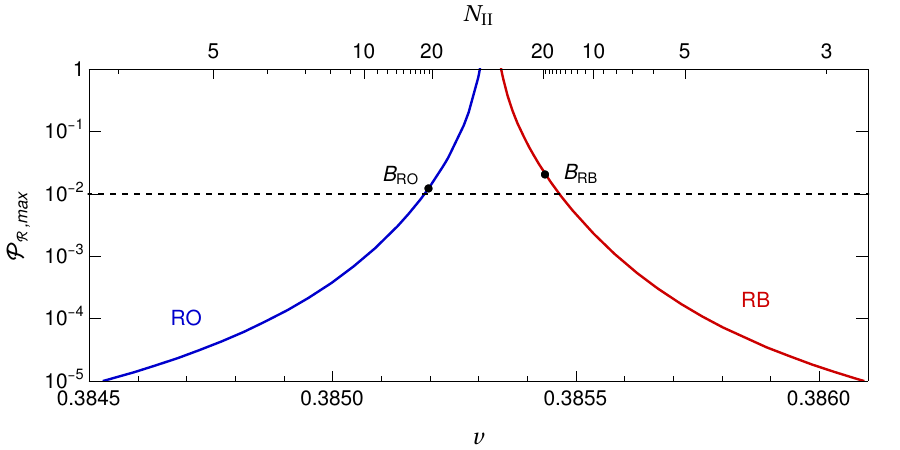}
    \caption{The maximal value of the power spectrum $\mathcal{P}_{\mathcal{R},\text{max}}$ as a function of $v$ for $\xi=0.01$. The upper $x$-axis denotes the corresponding duration of the second stage of inflation in $e$-folds. The inflaton stops at the hilltop for $v\approx 0.385325$. Roll-over (RO) and roll-back (RB) are denoted in red and blue, respectively. The black dots correspond to the benchmark point B in Table~\ref{tab:benchamrk_pts}.}
    \label{fig:PRmax}
\end{figure*}

%PBH-producing inflationary models typically suffer from fine-tuning (see~\cite{Cole:2023wyx} for a recent discussion). \R{[Extend!]}\R{ the need to set the model parameters with great accuracy to produce a high enough peak in the power spectrum.} 

PBH-producing inflationary models typically suffer from fine-tuning --- the model parameters must be set with great accuracy to produce a high enough peak in the power spectrum. There is no canonical measure of fine-tuning in models for PBHs (see, however, Ref.~\cite{Cole:2023wyx}) and we will characterize the level of fine-tuning by $-\log_{10}|v/v_c - 1|$ where $v$ is the vacuum expectation value corresponding to a $\mathcal{O}(10^{-2})$ peak in the curvature power spectrum and $v_c$ corresponds to the point at which a classical inflaton could be balanced exactly at the local maximum. This corresponds roughly to the number of significant digits $v$ needs to be tuned to obtain a sizable PBH abundance. For reference, in typical models for PBHs, %\eg, inflection-point models, 
the parameters of the potential must be tuned at the level of %3 or more 
5 to 8
significant digits~\cite{Cole:2023wyx}, with possibly the only exception being non-polynomonial inflection point inflation which seems to require tuning at the level of 3 significant digits~\cite{Cole:2023wyx}. We illustrate this for our model in Fig.~\ref{fig:PRmax}, where we display the peak value of $\PR$ as a function of the $v$ and $N_{\rm II}$. As $v$ approaches the critical value $v_c\approx 0.385325$, both $N_{\rm II}$ and $\mathcal{P}_{\mathcal{R},\rm max}$ approach infinity. However, the sensitivity is not exponential; we find
\be
    \mathcal{P}_{\mathcal{R},\rm max} \approx 5\times 10^{-16} |v/v_c - 1|^{-3.9}\,
\ee
when $N_{\rm II} < 20$, while the power-law behaviour is slightly milder for larger $N_{\rm II}$. All in all, to obtain $\mathcal{P}_{\mathcal{R},\rm max} \approx \mathcal{O}(10^{-2})$ we would need a fine-tuning of the order $v/v_c - 1 \approx 10^{-3.4}$, that is, at the level of the 3rd or 4th significant digit. This is less than in most inflection-point models~\cite{Cole:2023wyx}. In the same vein, Table~\ref{tab:benchamrk_pts} lists the tuned parameter values $v$ in example roll-over and roll-back cases with common digits colored in gray. These correspond to high peaks; for values in-between, the peaks are even higher. Tuning $v$ up to 4 digits is enough to guarantee abundant PBH production.

%%%%%%%%%%%%%%%%%%%%%%%%%%%%%%%%%%%%%%%%%%%%%%%%%%%%%%%%%%%%%%
\section{Discussion}
\label{sec:discussion}
%%%%%%%%%%%%%%%%%%%%%%%%%%%%%%%%%%%%%%%%%%%%%%%%%%%%%%%%%%%%%%

\subsection{Properties of the power spectrum}
To understand the shape of the power spectra of Fig.~\ref{fig:example_spectra}, we compare our results to an analytical approximation proposed in~\cite{Karam:2022nym}, which was almost exact for potentials built out of two downward-opening parabolas; the field starts by rolling down the side of one parabola in SR, then crosses the local minimum where the parabolas meet, and finally rolls over the local maximum of the other parabola in subsequent USR and CR phases. The $(aH)^{-2}z''/z$ term in the Mukhanov--Sasaki equation~\eqref{eq:SM_eom} consists of two plateaus, one for the SR phase and another for the dual USR/CR phases, with a delta function transition in-between. The result was a peaked, oscillating power spectrum, given by equation~(4.3) of~\cite{Karam:2022nym}:
\begin{equation} \label{eq:PR_approximation}
    \mathcal{P}_{\mathcal{R}}(k)  \approx \left|\theta(\mathcal{H}_c-k)\left[\frac{H^2}{2\pi\dot\varphi}\right]_{\mathcal{H}=k}
     + \frac{k \, p(k/\mathcal{H}_c,\lambda_2)}{k_{\rm CR}|p(k_{\rm CR}/k_c,\lambda_2)|} \left[\frac{\Gamma(\lambda_2)}{2^{3/2-\lambda_2}\pi^{3/2}}\frac{H^2}{\dot\varphi}\right]_{k=k_{\rm CR}}\right|^2 \, ,
\end{equation}
where the first term corresponds to SR and the second one to the USR and CR regimes. Here
\be
p(\kappa,\lambda_2) = e^{i\kappa}\left(\frac{\kappa}{2}\right)^{1-\lambda_2}\left(-J_{\lambda_2}(\kappa)+\left(i-\frac{1}{\kappa}\right)J_{\lambda_2+1}(\kappa)\right) \, ,
\quad
\lambda_2 = \frac{3}{2}\sqrt{1-\frac{4}{3}\eta_{V,\rm II}}\, ,
\ee
$k_{\rm CR}$ corresponds to some scale in the CR regime, $\mathcal{H}_c$ corresponds to the beginning of USR, $\mathcal{H}\equiv aH$, and $\eta_V \equiv V''/V$, with $\eta_{V,\rm II}$ evaluated at the potential's local maximum.

\begin{figure*}
    \centering
    \includegraphics[width=0.95\textwidth]{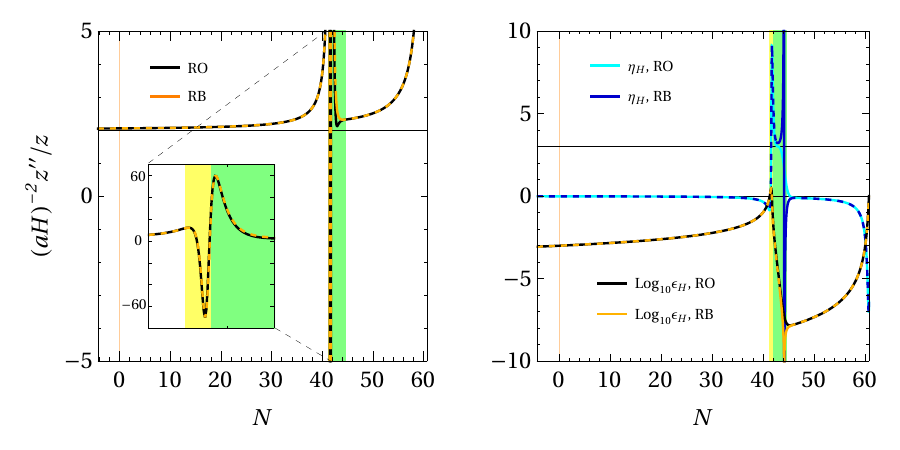}
    \caption{\emph{Left panel:} The behaviour of $(aH)^{-2}z''/z$ in the cases of roll-over (RO) and roll-back (RB), corresponding to point $B$ in Table~\ref{tab:benchamrk_pts}.  The yellow and green bands represent the temporarily halted inflation and USR, respectively. The black horizontal line is at $2$, corresponding to the value in SR. \emph{Right panel:} The behaviour of Hubble SR parameters $\epsilon_H$ and $\eta_H$ for roll-over and roll-back. In the case of roll-back, the field velocity becomes zero at the end of USR and changes sign as the field turns around near the origin. This is reflected in the behaviour of  $\eta_H$: it first goes to positive infinity and then to negative infinity as the field velocity changes sign.
    }
    \label{fig:zPrimePrimeAndEpsilonEta}
\end{figure*}

The setup of this paper is close to that of~\cite{Karam:2022nym}, therefore it makes sense to compare our power spectra to Eq.~\eqref{eq:PR_approximation}, with $\eta_{V,\rm II}=-4\xi$ by Eq.~\eqref{eq:eta0}. The comparison is made in Fig.~\ref{fig:example_spectra}. Both spectra exhibit oscillations followed by a CR plateau, and the plateau heights match, although our spectrum soon turns down to follow the SR approximation. Notably, the leading oscillation peak is much stronger in our case than in the approximation, completely overwhelming the rest of the spectrum. The difference is a consequence of the momentary halting of inflation, absent in the simplified model of Ref.~\cite{Karam:2022nym}. Fig.~\ref{fig:zPrimePrimeAndEpsilonEta} shows $(aH)^{-2}z''/z$ for our model, and it still exhibits the SR and USR/CR plateaus as in~\cite{Karam:2022nym}, but the behaviour in-between, when inflation momentarily ends, is complicated. In addition, some modes near the peak cross the Hubble radius multiple times: they exit in the initial phase of inflation, re-enter when inflation momentarily ceases, and exit again after inflation continues. To resolve these modes correctly, we must start tracking their behaviour before the first exit when they initially leave the Bunch--Davies vacuum. The details are discussed in Appendix~\ref{appendix:numerics}. The complicated behaviour strongly excites the Fourier modes, producing the peak.

Note that in~\cite{Karam:2022nym}, the oscillations in the spectrum were a consequence of the artificial sharp features in the approximation; in our case, the end of inflation naturally produces sharp features even with a smooth potential. The features arise from the high mass scale that dominates the time evolution when the field crosses the bottom of the potential. The spectral oscillations are analogous to excitations in preheating, where the field oscillates around its minimum after inflation and the oscillations induce particle production in periodic resonance bands, see \eg~\cite{Kofman:1994rk, Kofman:1997yn, Greene:1997fu}.

Even though~\eqref{eq:PR_approximation} fails to produce the strong peak, it qualitatively captures another feature of the spectrum: the presence or absence of a dip at the SR--USR boundary. Looking at~\eqref{eq:PR_approximation}, the field velocities $\dot \phi$ in the two terms have the same sign in roll-over, but different signs in roll-back, since there the field changes direction before entering the CR regime. The roll-over case then experiences destructive interference from the two terms, producing a dip, while such behaviour is absent in roll-back, and thus no dip is formed. A dip could be detectable in future $\mu$-distortion experiments~\cite{Ozsoy:2021pws} or 21 cm observations~\cite{Balaji:2022zur}. In light of our results, such observations could distinguish roll-over and roll-back-type models from each other. %However, loop corrections have been shown to significantly reduce the depth of the dip~\cite{Franciolini:2023lgy}.

Recently, the suggestion that one-loop corrections may rule out single-field inflationary scenarios for PBHs~\cite{Kristiano:2022maq} has sparked an active debate in the literature~\cite{Riotto:2023hoz, Riotto:2023gpm, Firouzjahi:2023ahg, Firouzjahi:2023aum, Kristiano:2022maq, Kristiano:2023scm, Choudhury:2023vuj, Choudhury:2023jlt, Choudhury:2023rks, Franciolini:2023lgy}. The recent study~\cite{Franciolini:2023lgy} showed that although appreciable one-loop corrections can be generated, they do not conflict with the CMB in models producing a significant PBH abundance. The general claim is that, assuming single-field inflation, an enhanced peak in the power spectrum at small scales resulting from an extended period of USR will induce large corrections at the CMB scales. The duration of USR is related to the heights of power spectra during the first and second inflationary SR/CR periods. In the double-well scenarios, the duration of the USR period can be significantly shorter than in the quasi-inflection point ones, suggesting that the one-loop corrections may be diminished when compared to the quasi-inflection point scenarios. However, a dedicated study would be required to confirm this claim.\footnote{Soon after the submission of the initial version of the paper, it was shown that including boundary terms leads to a volume-suppressed one-loop correction.}

In addition to the case of a single roll-over or roll-back, our scenario can be expanded to multiple roll-overs and roll-backs in a landscape of potential minima. One can envisage, for example, the field rolling over both of the minima of a double-well potential relatively unhindered, and entering a roll-back phase on the subsequent slope on the other side of the potential, or even continuing from there and undergoing a roll-over phase on its way back, when crossing the potential maximum a second time. A case like this was considered in \cite{Saito:2008em}. In principle, it is possible to realize multiple USR--CR periods this way, producing multiple peaks in the power spectrum. Indeed, multiple roll-back-like phases happen in usual reheating when the field oscillates around a minimum, switching direction periodically. Usually, no discernable amount of inflation takes place between these; stretching the produced perturbations to super-Hubble scales takes additional tuning.

Finally, the strong power spectrum peak will induce a stochastic GW background (SGWB) at the second order of perturbation theory so that all the scenarios mentioned above can be tested by future GW experiments~\cite{Saito:2008jc, Assadullahi:2009jc, Bugaev:2010bb, Alabidi:2012ex, Inomata:2016rbd, Orlofsky:2016vbd, Espinosa:2018eve, Inomata:2018epa, Byrnes:2018txb, Cai:2018dig, Bartolo:2018rku, Wang:2019kaf, Yuan:2019udt, Chen:2019xse, Karam:2022nym}. LISA~\cite{LISA:2017pwj} and AEDGE~\cite{AEDGE:2019nxb, Badurina:2021rgt} may probe the SGWB generated in asteroid mass PBH scenarios, while pulsar timing arrays (PTAs) can test the curvature fluctuations associated with solar mass PBHs. In fact, the stochastic common-spectrum process recently detected by pulsar timing experiments~\cite{NANOGrav:2020bcs, Goncharov:2021oub, Chen:2021rqp, Antoniadis:2022pcn} may be interpreted as such a primordial SGWB induced by large scalar perturbations~\cite{Vaskonen:2020lbd, DeLuca:2020agl, Kohri:2020qqd, Domenech:2020ers, Dandoy:2023jot}.

%\cite{Yokoyama:1998pt, Saito:2008em, Bugaev:2008bi, Briaud:2023eae}

\subsection{Comparison to previous studies}

Early studies of PBH production from double-well potentials include \cite{Yokoyama:1998pt, Saito:2008em, Bugaev:2008bi}. They obtained a similar shape for the power spectrum, with a strong peak followed by rapid oscillations and a plateau. However, the models of \cite{Yokoyama:1998pt, Saito:2008em, Bugaev:2008bi} do not reproduce the CMB observations~\eqref{eq:CMB_observables}. We have presented a model where this is possible and concentrated on PBHs in the asteroid mass window. We have also discussed the general properties of such models based on recent developments in inflationary model building and studied the power spectrum of the roll-back case, absent from \cite{Yokoyama:1998pt, Saito:2008em, Bugaev:2008bi}.

A model similar to ours with roll-back was constructed in~\cite{Fu:2020lob} by stitching together two different potentials on two sides of a single minimum. Strong, PBH-producing perturbations were produced after the inflaton had crossed the minimum and turned back. In \cite{Fu:2022ypp}, a more physical model was constructed with a smooth potential and a non-canonical kinetic term featuring a double pole. The potential in terms of the canonical field has two plateaus at different heights with a minimum in-between. Again, strong perturbations were generated, but the model did not provide the correct CMB predictions.
Both \cite{Fu:2020lob} and \cite{Fu:2022ypp} obtain power spectra with no dip before the peak, consistent with our results. They also obtain high-frequency oscillations in the tail of the peak. However, in our case, see Fig.~\ref{fig:example_spectra}, the leading oscillation is much more pronounced than in~\cite{Fu:2020lob, Fu:2022ypp}, and the subsequent oscillations die down faster and do not exhibit similar deep dips. 

Such slow decay of the oscillations may be related to the sharp features in the potential, for instance, in~\cite{Fu:2020lob}, the potential was constructed by gluing together two smooth potentials at the minimum. However, Ref.~\cite{Fu:2022ypp} does find the same effect with a smooth potential. One can draw parallels with inflation fragmentation during preheating, for which it has been shown that the Floquet exponents, which control mode growth during a half-oscillation, are bounded by a power-law in $k$ if the Mukhanov-Sasaki equation contains discontinuities in derivatives of the potential. On the other hand, an exponential damping in $k$ is observed for a smooth potential~\cite{Koivunen:2022mem}. The damping of mode growth in $k$ will imply the damping of spectral oscillations. Similar smoothing of the potential has been noted in the context of toy models for PBHs~\cite{Cole:2022xqc, Karam:2022nym}. Despite the hints provided by these studies, further research is required to establish the exact nature of the spectral oscillations.

The recent paper~\cite{Briaud:2023eae} discusses a setup similar to ours, concentrating on the crossing of the local maximum of a double-well potential which is associated with PBH formation but not the CMB. The authors go beyond our linear analysis by using stochastic inflation allowing them to study non-Gaussianities and their effect on PBH formation. However, the linear perturbation spectrum is still needed as an input for the stochastic calculation. The authors of~\cite{Briaud:2023eae} use an analytic expression for the spectrum, capturing the (approximately) constant slope near the leading peak in our Fig.~\ref{fig:example_spectra}. Our analysis can complement the approach of~\cite{Briaud:2023eae} by providing a more accurately computed power spectrum, including the highest peak and the cutoff at low $k$, which in our model is highly enhanced and may thus modify the stochastic results. We leave the stochastic analysis of our model for future work.

%OLD: The recent paper~\cite{Briaud:2023eae} discusses a setup similar to ours with a double-well potential. The authors go beyond our linear analysis by using stochastic inflation, a computational method that presents corrections to the Gaussian PBH abundance. However, the linear perturbation spectrum is still needed as an input for the stochastic calculation. The authors of~\cite{Briaud:2023eae} use a simple expression for the spectrum, capturing the (approximately) constant slope after the leading peak in our Fig.~\ref{fig:example_spectra}, but not the peak itself or the cutoff at lower $k$. The model of~\cite{Briaud:2023eae} also restricts their analysis to very light PBHs, and they do not try to fit the CMB. We leave the stochastic analysis of our model for future work.

%%%%%%%%%%%%%%%%%%%%%%%%%%%%%%%%%%%%%%%%%%%%%%%%%%%%%%%%%%%%%%
\section{Conclusions}
\label{sec:concusions}
%%%%%%%%%%%%%%%%%%%%%%%%%%%%%%%%%%%%%%%%%%%%%%%%%%%%%%%%%%%%%%

We examined the formation of large peaks in the curvature power spectrum from inflation driven by double-well potentials with a $\mathbb{Z}_2$ symmetry. After the initial inflationary phase, the rolling inflaton crosses one of the minima and slows down rapidly around the local maximum, creating a peak in the curvature power spectrum. Then the inflaton either has enough energy to cross the hilltop, or it does not have enough energy to reach the origin and it turns around. In both cases, an ultra-slow-roll phase ensues followed by a slow-roll/constant-roll phase dual to it. The main difference between rolling over or turning around is that the roll-back spectrum displays no dip before the peak, which may be detected by future $\mu$-distortion experiments or 21 cm observations.

We constructed an explicit double-well scenario that is consistent with the latest CMB observations and capable of generating sufficiently large curvature perturbations for primordial black holes. The peaks in the curvature power spectra are more prominent than those in the commonly studied quasi-inflection point scenarios and typically require less tuning of the model parameters. The enhanced peak arises from the temporary cessation of inflation as the field traverses the minimum of the potential. Our results suggest that double-well potentials offer a promising avenue for further investigation and could provide important insights into primordial black hole phenomenology.

%Finally, we note that double-well scenarios do not necessarily rely on a notable period of ultra-slow-roll to generate a significant primordial black hole abundance. Thus we expect them to provide an example of single-field inflation for PBHs for which one-loop corrections are suppressed when compared to quasi-inflection point scenarios.

%-------------------------------------------------------------------------------
\acknowledgments
%-------------------------------------------------------------------------------

We thank Antonio Iovino for useful discussions. This work was supported by the Estonian Research Council grants PSG869, PRG1055, PRG1677, PSG761 and by the EU through the European Regional Development Fund CoE program TK133 ``The Dark Side of the Universe".

\appendix

\section{Solving primordial perturbations}\label{appendix:numerics}
The evolution of scalar curvature perturbation $\mathcal{R}_c$ obeys the Mukhanov--Sasaki equation. It is traditionally cast into a convenient form  using the Mukhanov field $u\equiv z\mathcal{R}_c$, where
\be
    z\equiv \frac{a\dot\varphi}{H} \, .
\ee
The Fourier modes of the Mukhanov variable evolve according to the Mukhanov--Sasaki equation~\cite{Mukhanov:1985rz, Sasaki:1986hm, Mukhanov:1990me}:
\be \label{eq:u_eom}
    u_k''+\left(k^2-\frac{z''}{z}\right)u_k=0 \, ,
\ee
where the prime denotes differentiation with respect to conformal time, $d\tau = dt/a$. 

In the sub-Hubble regime, $aH\ll k$, the curvature of space-time is negligible and the space-time is essentially Minkowski. A given mode starts its evolution deep inside the horizon. This leads to Bunch--Davies initial conditions
\be
    u'_k=-ik u_k \, , \quad |u_k|=\frac{1}{\sqrt{2k}}
\ee
at $\tau\to-\infty$. The power spectrum $\mathcal{P}_{\mathcal{R}}$ is given by
\be
    \mathcal{P}_{\mathcal{R}} = \frac{k^3}{2\pi^2}\frac{|u_k|^2}{z^2} \, ,
\ee
evaluated at late times.

We solve the Mukhanov--Sasaki equation numerically. For the numerical implementation, it is convenient to introduce a new variable, closely related to the Mukhanov variable $u$~\cite{Rasanen:2018fom},
\be\label{eq:newVariable1}
    g_k\equiv \frac{u_k e^{ik\tau}}{z} \, .
\ee
The advantage of this variable is that the multiplication with $e^{ik\tau}$ cancels the rapid early oscillations in $u_k$. The $g_k$ is also not subject to as large changes as $u_k$ in its order of magnitude due to the division by $z$. The Bunch--Davies vacuum condition in terms of $g_k$ is
\be \label{eq:g_Bunch_Davies}
    \dot g_k = -\frac{\dot z}{z}g_k \, ,\quad |g_k|=\frac{1}{\sqrt{2k}|z|} \, ,
\ee
and for mode $k$ we start our numerical evolution at time $t$, defined by $a(t)H(t)=k/100$. In our case with inflation stopping momentarily, we need to take special care in applying the initial condition, since some scales re-enter the Hubble radius during the brief period with $\epsh > 1$ and exit again when inflation continues. To make sure we capture the full non-trivial time evolution of a mode, we start tracking each of them before their first Hubble exit.

The Mukhanov--Sasaki equation in terms of $g_k$ is
\be \label{eq:g_eom}
    \ddot g_k +\left(H+2\frac{\dot z}{z}-\frac{2ik}{a}\right)\dot g_k
-\frac{2ik}{a}\frac{\dot z}{z}g_k=0 \, ,
\ee
where the dot denotes differentiation with respect to the cosmic time $t$. The conformal time grows exponentially with cosmic time and hence the dynamical range of cosmic time is smaller than that of conformal time, therefore the use of cosmic time allows for an easier numerical treatment.

As is well known, in SR inflation the curvature perturbations---and thus $g_k$---freeze to a constant value after they exit the Hubble radius. When the SR conditions are broken, such as during our non-inflating period and the subsequent phase when the inflaton climbs up towards the potential maximum, this is not necessarily the case \cite{Kinney:2005vj, Rasanen:2018fom}---indeed, the power spectrum enhancement of Fig.~\ref{fig:example_spectra} is a consequence of the super-Hubble growth of the corresponding modes. The spectrum only freezes in the final SR phase after the roll-over or roll-back has happened. To make sure we capture the final value of the spectrum, we track the modes all the way to the end of inflation.

We use the variable $g_k$ and equations \eqref{eq:g_Bunch_Davies}, \eqref{eq:g_eom} for the roll-over case where the inflaton field rolls over the local maximum at the origin. However, $g_k$ is not suitable for the roll-back case where the field stops near the local maximum at the origin and starts to roll back, since $z$ becomes zero at the turning point and $g_k$ is not defined. This is a gauge issue: the comoving gauge is momentarily not defined, and neither is $\mathcal{R}$. However, the Mukhanov variable $u_k$ turns out to behave well even in this limit (in particular, $(aH)^{-2}z''/z$ in~\eqref{eq:u_eom} behaves regularly); it is related to the field perturbation in the spatially flat gauge. In the spirit of \eqref{eq:newVariable1}, we introduce the new variable
\be\label{eq:newVariable2}
    h_k=\frac{u_k e^{ik\tau}}{a} \, ,
\ee
which stays well-defined throughout the roll-back case. The Bunch--Davies vacuum initial condition in this case is
\be
    \dot h_k = -\frac{\dot a}{a} h_k \, , \quad |h_k|=\frac{1}{\sqrt{2k}|a|} \, ,
\ee
and the Mukhanov--Sasaki equation is
\be
    \ddot h_k +H\left(3-\frac{2ik}{aH}\right)\dot h_k
+H^2\left(2-\epsilon_H-2i\left(\frac{k}{aH}\right)-\frac{H\dot z+\ddot z}{H^2 z}\right)h_k=0 \, .
\ee
The variable $h_k$ is suitable for both roll-over and roll-back, but is slightly slower to solve numerically. In the numerical analysis we use the variable $g_k$ for roll-over; in the case of roll-back we use the variable $g_k$ from the beginning until the re-start of inflation and then switch to $h_k$, before the field turns around.

\bibliography{PBHinflation}

\end{document}